\documentclass[10pt,aps,pre,twocolumn,showpacs,superscriptaddress,amsmath,amssymb,longbibliography]{revtex4-1}
\usepackage{amsmath}
\usepackage{graphicx}
\usepackage{xcolor}

\usepackage{todonotes}

\begin{document}
	
\title{Mechanical equilibrium of aggregates of dielectric spheres}

\author{A. F. V. Matias}
\email{afmatias@alunos.fc.ul.pt}
\affiliation{Departamento de F\'{\i}sica, Faculdade de Ci\^{e}ncias, Universidade de Lisboa, 1749-016 Lisboa, Portugal}
\affiliation{Centro de F\'{i}sica Te\'{o}rica e Computacional, Universidade de Lisboa, 1749-016 Lisboa, Portugal}

\author{T. Shinbrot}
\email{shinbrot@rutgers.edu}
\affiliation{Department of Biomedical Engineering, Rutgers University, Piscataway, New Jersey, 08854, USA}    
	
\author{N. A. M. Ara\'ujo}
\email{nmaraujo@fc.ul.pt}
\affiliation{Departamento de F\'{\i}sica, Faculdade de Ci\^{e}ncias, Universidade de Lisboa, 1749-016 Lisboa, Portugal}
\affiliation{Centro de F\'{i}sica Te\'{o}rica e Computacional, Universidade de Lisboa, 1749-016 Lisboa, Portugal}

\begin{abstract}
	Industrial as well as natural aggregation of fine particles is believed to be associated with electrostatics.  Yet like charges repel, so it is unclear how similarly treated particles aggregate.  To resolve this apparent contradiction, we analyze conditions necessary to hold aggregates together with electrostatic forces.  We find that aggregates of particles charged with the same sign can be held together due to dielectric polarization, we evaluate the effect of aggregate size, and we briefly summarize consequences for practical aggregation.
\end{abstract}

\maketitle


\section{Introduction}

Electrostatic attraction is central to both industrial and natural aggregation.  Industrially, this leads to profound problems, for example producing variations in active ingredient concentrations by as much as 100\% in common pharmaceuticals \cite{Mehrotra2007}. Likewise sticking of thick layers of polymers to fluidized bed risers \cite{Hendrickson2006,Tian2015} forces manufacturing shut-downs to scour the risers.  Particle aggregation also contributes to material inhomogeneities that produce a documented 50\% rejection rate in manufactured ceramics \cite{Bowen1983}.  In nature, planets emerge in protoplanetary disks from the aggregation of dust particles.  In the size range below about 10 $\mu$m, particles can stick due to van der Waals forces \cite{doi:10.1146/annurev.astro.46.060407.145152}, and at sizes above centimeters, gravity can account for aggregation \cite{BODENHEIMER1986391}.  Between these sizes, it is unclear what produces observed aggregates \cite{Zsom2010,kelling2014experimental,Kruss2017}.

Experiments \cite{Spahn2015,article}, analytic calculations \cite{Nakajima1999,Jones1979}, and numerical simulations \cite{Qin2016,Feng2000,Gan2015} suggest that electrostatic interactions could produce aggregation in this intermediate range, however like charged grains should repel, and so electrostatic attraction requires charge heterogeneity.  It is currently unclear how grains acquire sufficiently dissimilar charges to avoid Coulomb repulsion, and how dissimilar these charges must be to account for the formation of multiparticle aggregates.

Recent computations by Feng \cite{Feng2000} provide insight into this problem by demonstrating that two dielectric particles with identical sign charges can attract if the magnitude of charges on the particles differ significantly, and the particle permittivity is sufficiently high.  The mechanism explored in Ref \cite{Feng2000} is that induced dipole-dipole attraction can overcome Coulomb repulsion if the former is large and the latter is small.  

In the present work, we investigate whether this mechanism can account for multi-particle aggregates, and we identify conditions under which these aggregates can form.  The manuscript is organized as follows. In the next section, we introduce the model. We discuss the results in Sec. \ref{sec:results} and present some final remarks in Sec. \ref{sec:discussion}.


\section{Model}
\label{sec:theory}

\begin{figure}[t]
	\centering
	\includegraphics[width=0.95\linewidth]{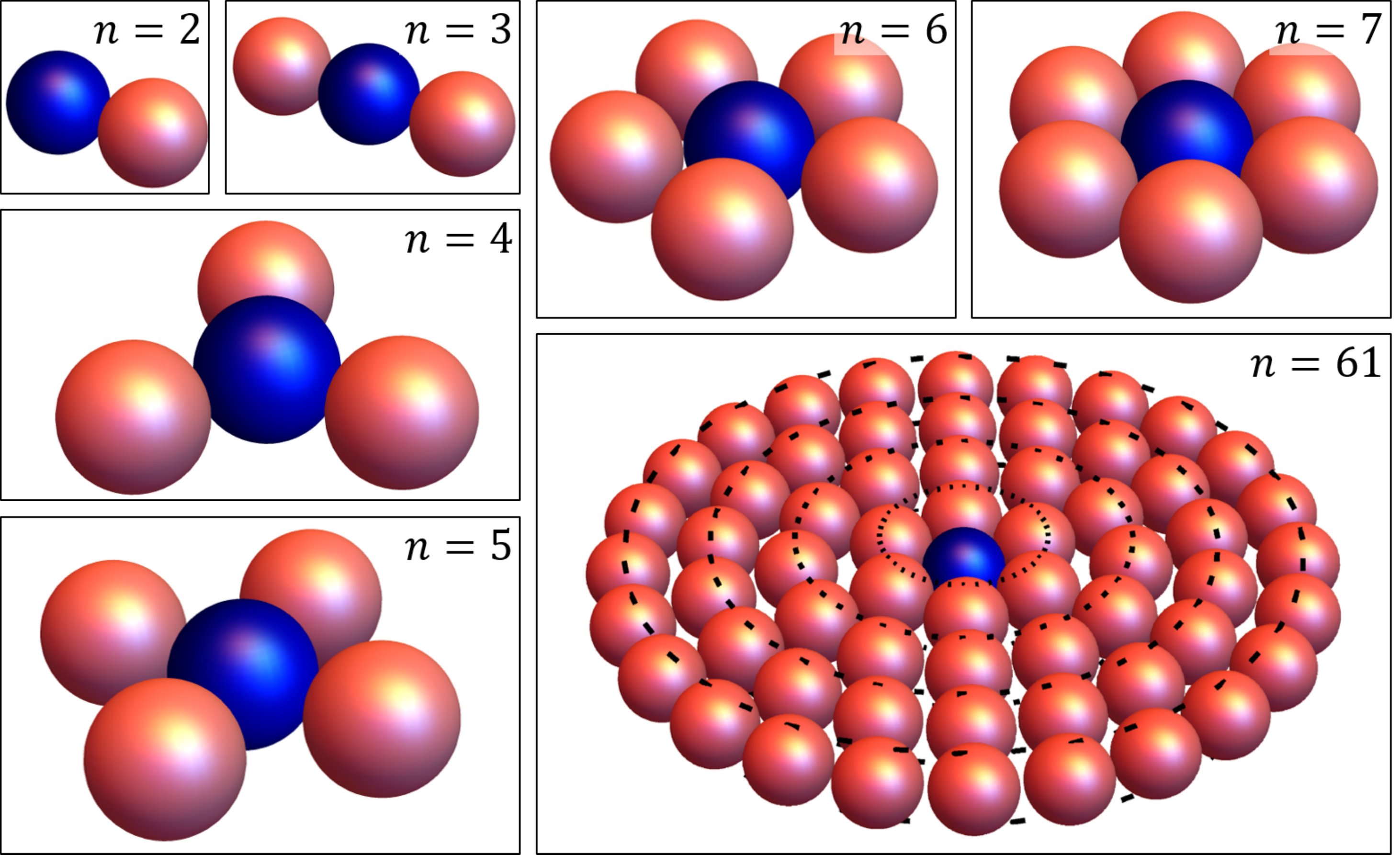}
	\caption{Schematics of the aggregates considered here, for numbers, $ n $, of particles indicated. Dielectric spheres have the same radius $ R $ and dielectric constant $ \varepsilon_p $. The charge of the central sphere (blue online) is $ Q_c $ and the surrounding spheres (red online) have fixed charge $ Q_0 $ each.}
	\label{fig:example_system}
\end{figure}

To systematically investigate electrostatic forces on aggregates as a function of their size, we consider spherical dielectric particles with the same radius $ R $ and permittivity $ \varepsilon_p $, surrounded by a medium of permittivity $ \varepsilon_0 $, as illustrated in Fig. \ref{fig:example_system}. We evaluate increasing numbers, $n$, of particles as shown, where for $n\leq 7$, the particles are placed symmetrically around a central particle. For $n >7$ complete shells of particles are layered concentrically as shown.  We fix the charge of the surrounding particles at $ Q_0 $ each, and define the single central  particle to have charge $ Q_c $.  Here, we consider that the charges are distributed uniformly on the surface of each particle and we evaluate the force on each particle by solving the Poisson equation \cite{GriffithsElectrodynamics,Jackson1962} for the electric field $ \vec{E} $ 
\begin{equation}
\nabla\cdot \left[\varepsilon(\vec{x})\vec{E}(\vec{x})\right] = -\nabla\cdot \left[\varepsilon(\vec{x})\nabla V(\vec{x})\right] = \rho(\vec{x}) \mathrm{,}
\label{eq:poisson}
\end{equation}
where $ V $ is the electrostatic potential, $ \rho $ is the charge density, and $ \varepsilon $ is the permittivity. 

All variables are made dimensionless by measuring distances in units of $ R $, charges in units of $ Q_0 $, dielectric constants in units of $ \varepsilon_0 $, and electrostatic forces in units of $ Q_0^2/\left[4\pi\varepsilon_0 (2R)^2\right] $. Accordingly, all surrounding particles are of unitary charge and the central particle has charge $ Q=Q_c/Q_0 $.  Thus
\begin{equation}
\varepsilon(\vec{x})\nabla\cdot\vec{E}(\vec{x})=\rho(\vec{x})-\nabla\varepsilon(\vec{x})\cdot\vec{E}(\vec{x}) =
\rho(\vec{x}) + \rho_p(\vec{x})
\mathrm{,}
\label{eq:polarization}
\end{equation}
where $ \rho_p(\vec{x})=-\nabla\varepsilon(\vec{x})\cdot\vec{E}(\vec{x}) $ is the polarization charge density. For particle permittivity $\varepsilon_p$, equal to vacuum permittivity, $\varepsilon_0$, the polarization charge density is zero. But when $\varepsilon_p$ is different from $\varepsilon_0$, polarization charges arise that alter the electric field, and consequently forces between particles.

At the boundary, $ \partial \Omega $, of each particle we impose
\begin{equation}
\begin{aligned}
&V_{in}(\vec{x}) = V_{out}(\vec{x})
\\
&\vec{n}\cdot\left(\kappa\nabla V_{in}(\vec{x}) - \nabla V_{out}(\vec{x}) \right) = \sigma
\end{aligned}
\quad
\text{for }\vec{x} \in \partial \Omega \mathrm{,}
\label{eq:boundary_particle}
\end{equation}
where $ \vec{n} $ is the unitary vector normal to the particle surface, $ \sigma $ is the surface charge density of the particle, the subscripts $ in $ and $ out $ denote the potential inside and outside the particle, and $ \kappa=\varepsilon_p/\varepsilon_0 $ is the dielectric constant with respect to $\varepsilon_0$. Notice that when $ \kappa=1 $ the permittivity of the medium equals that of the particles, and $ \rho_p=0 $.

Far from the particles the electrostatic potential is dominated by the monopole term. Thus we consider that at the asymptotic boundary $ S_\mathrm{asymp} $, defined as a spherical shell of radius $ R_\mathrm{max}=20R $, the electrostatic potential is
\begin{equation}
V(\vec{x}) = \sum_i \frac{q_i}{\left| \vec{x}_i-\vec{x} \right|}
\quad
\text{for }\vec{x} \in S_\mathrm{asymp}\mathrm{,}
\label{eq:pot}
\end{equation}
where we sum over all particles $ i $ with charge $ q_i $ and center at $ \vec{x}_i $.

The force that acts on each particle is obtained by integrating the Maxwell stress tensor \cite{GriffithsElectrodynamics} over its surface $\partial\Omega$, \textit{i.e.}
\begin{equation}
\vec{F}=\iint_{\partial \Omega} \left[ \vec{E}\otimes\vec{E} - \frac{1}{2}\left(\vec{E}\cdot\vec{E}\right)\textbf{I}\right]\vec{n} dS
\mathrm{,}
\label{eq:force_maxwell}
\end{equation}
where $ \otimes $ is the dyadic product and $ \textbf{I} $ is the identity matrix.

\begin{figure}[t]
	\centering
	\includegraphics[width=\linewidth]{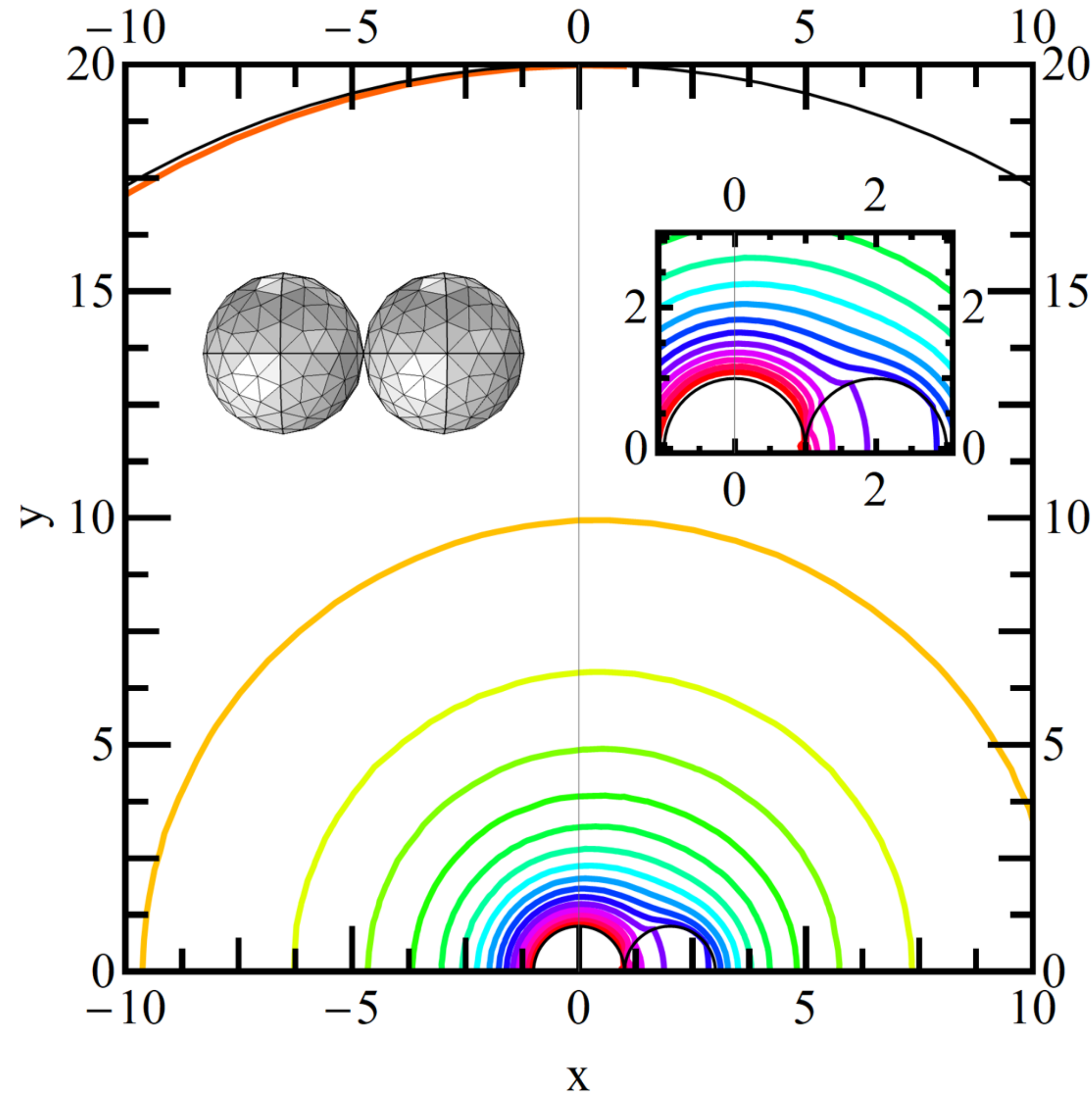}
	\caption{Equipotential lines for the case with $n=2$ and $\kappa=8$, the particle in the left-hand side has charge five and the other one has unitary charge. The equipotential values are in increments of $0.3$ from $0.3$ to $4.8$. The insets contain the mesh used on the boundary of the particles and a detail of the potential near the particles.}
	\label{fig:potential}
\end{figure}

Following Ref. \cite{Nakajima1999}, the radial component of the force acting on a surrounding particle due to the central one is given by
\begin{equation}
\begin{aligned}
F_r&=\frac{\kappa+2}{\kappa-1}\frac{1}{R}a_1+
\\
&+\frac{1}{\kappa-1}\sum_{n=1}^{\infty}\big[(n+1)(\kappa+1)+1\big]a_n a_{n+1}
\end{aligned}
\label{eq:force_analytical}
\end{equation}
where $a_i$ represents an infinite sum that is linear with respect to Q. Thus, the radial component of the electrostatic force $ F_r $ acting on a surrounding particle relates to the charge of the central particle according to a second order polynomial \cite{Jones1995}
\begin{equation}
F_r(Q) = \alpha Q^2 - \beta Q + \gamma \mathrm{,}
\label{eq:force_parabola}
\end{equation}
where the constant $\beta$ define the Coulomb interaction (first term in Eq. \eqref{eq:force_analytical}), and the constants $\alpha$ and $\gamma$ define the force due to particle polarization. Here the coefficients of Eq. \eqref{eq:force_parabola} are determined numerically by least square fitting the data obtained for different charge values in the range $ Q\in[-10,+10] $.

We consider that the aggregate is in mechanical equilibrium when the radial component of the net electrostatic force in the outer (red online) particles point toward the center. We assume that particles are rigid and not deformable. Thus the limits of the mechanical equilibrium correspond to the zeros of Eq. \eqref{eq:force_parabola}, given by
\begin{equation}
Q=\frac{\beta\pm\sqrt{\beta^2-4\alpha\gamma}}{2\alpha}\mathrm{.}
\label{eq:zeros}
\end{equation}
Note that the forces acting on the particles may change the configuration of the aggregate, even without separating the particles. For simplicity we we neglect particle rearrangement.

\begin{figure}[t]
	\centering
	\includegraphics[width=\linewidth]{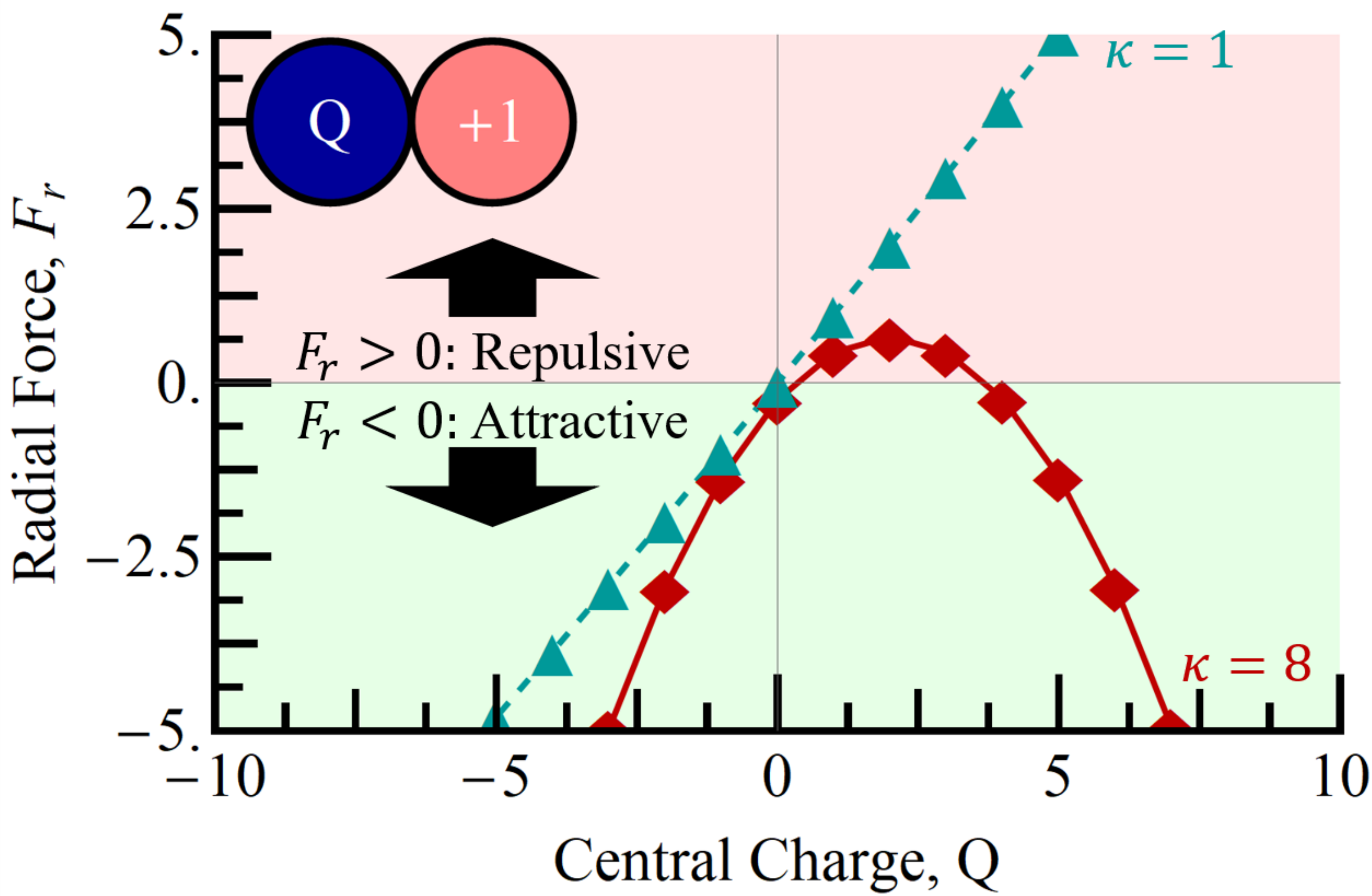}
	\caption{Radial component of the electrostatic force $F_r$ acting on the right-hand particle (of unitary charge) as function of charge $Q$ of the left-hand particle, for $ \kappa=1 $ (triangles, red online), and $ \kappa=8 $ (diamonds, green online). The lower half-plane defines attractive forces between the pair of particles \textit{i.e.}, $F_r<0$ (shaded green online).}
	\label{fig:Feng_simple}
\end{figure}

The spatial dependence of the electrostatic potential, inside a sphere with radius $20R$, and the electrostatic forces are calculated with the finite element method using COMSOL Multiphysics\textregistered~software \cite{COMSOL,strang2007computational}. The system is divided into a mesh of tetrahedral elements whose size is varied across the system according to accuracy needs. Inside the particles, where the accuracy needs to be the highest to calculate the electrostatic forces accurately, the elements are the smallest: their size is smaller than $ 1.1R $. Outside the particles the elements get progressively larger. On the boundary, where less a accurate field is required, the element size is larger than $ 0.36R $. The mesh on the surface of two contacting particles is represented in the inset to Fig. \ref{fig:potential}. The potential in each element, with position $(x,y,z$), is described by a second order polynomial
\begin{equation}
\begin{aligned}
	V(x,y,z)&=a_1 + a_2x + a_3y + a_4z + a_5xy +
    \\
	&+ a_6xz + a_7yz + a_8x^2 + a_9y^2 + a_{10}z^2
\end{aligned}
\mathrm{,}
\end{equation}
meaning there are 10 coefficients $a_i$ per element to be determined. To solve the linear system associated with the mesh and determine the coefficients, the MUMPS \cite{AMESTOY2006136} solver was used. A solution to the potential is found when the error estimate, related to the residue of the linear system according to Ref. \cite{comsolreference}, is smaller than $10^{-3}$.


\section{Results}
\label{sec:results}

Figure \ref{fig:potential} shows the equipotential lines for two touching particles, the particle in the left-hand side has charge five and the other one has unitary charge, with $\kappa=8$ (\textit{i.e.} $\varepsilon_p=8 \varepsilon_0$). Due to the existence of polarization charges, the electrostatic potential near the particles is distorted, as can be seen in the inset. As we move away from the particles the electrostatic potential gets less deformed since the polarization charges became relevant than the charges of the particles.

Let us start the study of the electrostatic forces with the case of two aggregated particles, as studied in detail by Feng in Ref. \cite{Feng2000}. Figure \ref{fig:Feng_simple} shows the radial component of the electrostatic force acting on the right-hand particle shown in the inset as a function of the charge $Q$, on the left-hand particle.  Two dielectric constants are shown, $\kappa=1$ and $\kappa=8$. For $\kappa=1$, the interaction between particles follows Coulomb's law
\begin{equation}
F_r=\frac{Q}{4\pi\varepsilon_0(2R)^2} \mathrm{,}
\end{equation}
so the attractive force between the particles scales linearly with $Q$ --- \textit{i.e.} $\alpha=\gamma=0$ in Eq. \eqref{eq:force_parabola}. In this case, the aggregate coheres only if the particles have charges of opposite sign. By contrast, given the appropriate conditions, for $\kappa=8$ and $ \rho_p\neq0 $, the aggregate coheres for particles with like-charges. In fact, $ \rho_p $ increases with the strength of the electric field and, consequently, for sufficiently large $ Q $, polarization charges dominate the electrostatic interaction and even same-sign particles can attract, as can be seen in Fig. \ref{fig:Feng_simple} for $Q>3.7$. Notice that the aggregate is cohesive for $Q=0$: this is the case of dielectrophoresis on a neutral particle.

\begin{figure}[t]
	\centering
	\includegraphics[width=\linewidth]{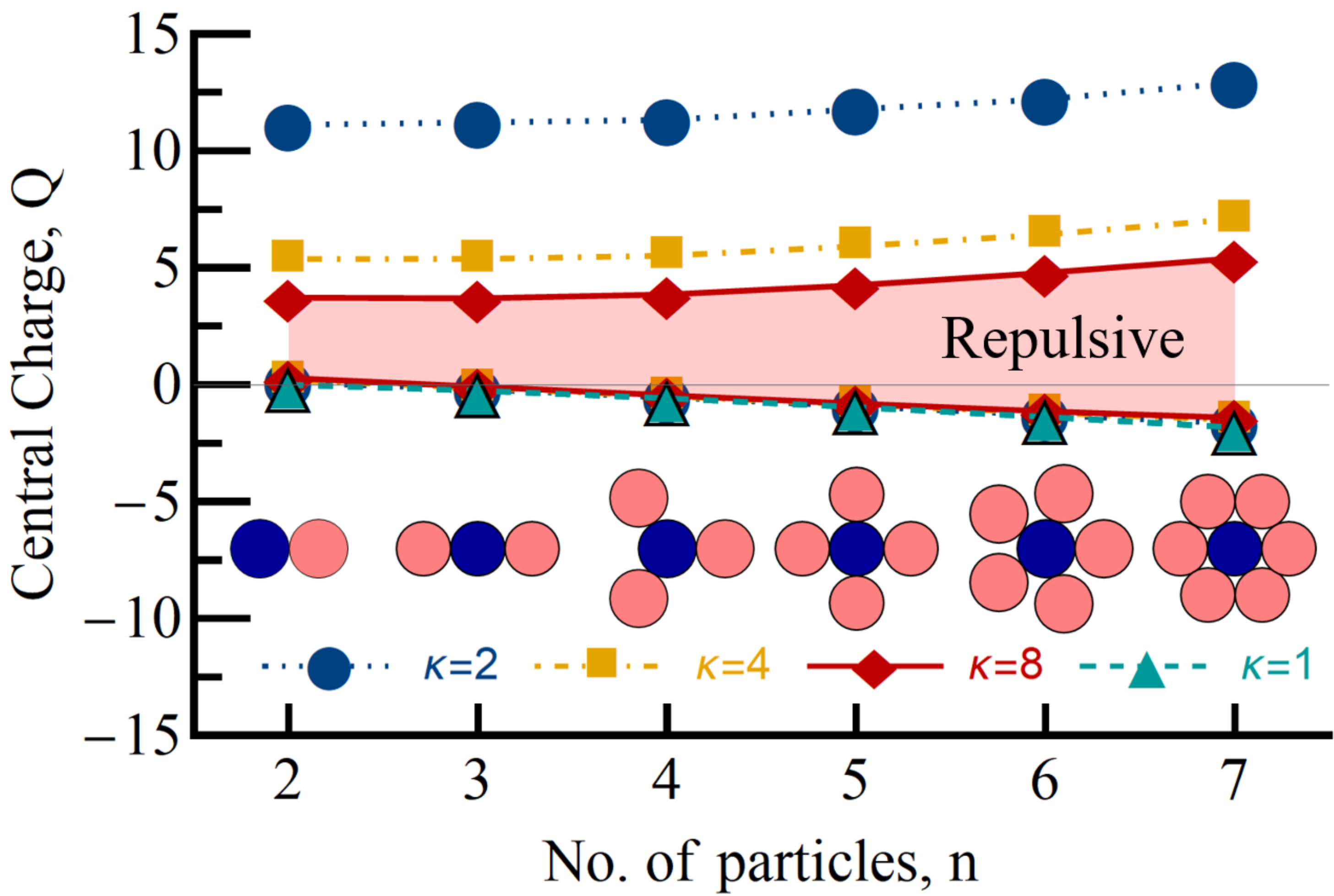}
	\caption{Plot of the limits for mechanical equilibrium, corresponding to the zeros of Eq. \eqref{eq:force_parabola}, as function of the size, $n$, of aggregates for different dielectric constants (values shown to the right of each curve). Particle arrangements shown beneath each value of $n$.  In the region between the curves, shaded for $ \kappa=8 $, the outermost particles experience a net outward radial force; outside the curves, all particles are radially attracted.}
	\label{fig:limits_short}
\end{figure}

Figure \ref{fig:limits_short} shows the limits of cohesion of aggregates of up to seven particles for several values of dielectric constant. Cohesion (\textit{i.e.}, all particles are attracted) is defined by two limits; in the region between the curves, shaded for $ \kappa=8 $, the outermost particles experience net outward radial force.  Due to the aggregate symmetry, the radial component of the force is the same for all of the outermost particles. For the lower limit, $ Q $ is negative, (except for $ n=2 $ which we have already discussed), thus the surrounding particles are attracted to the central one due to a combination of Coulomb and polarization forces. Note that above the upper curves, all particles have the same charge sign, thus polarization is responsible for cohesion of the aggregate. By contrast, since polarization grows with the dielectric constant, the minimum value of $Q$ for which the aggregates are cohesive decreases with $\kappa$.

\begin{figure}[t]
	\centering
	\includegraphics[width=\linewidth]{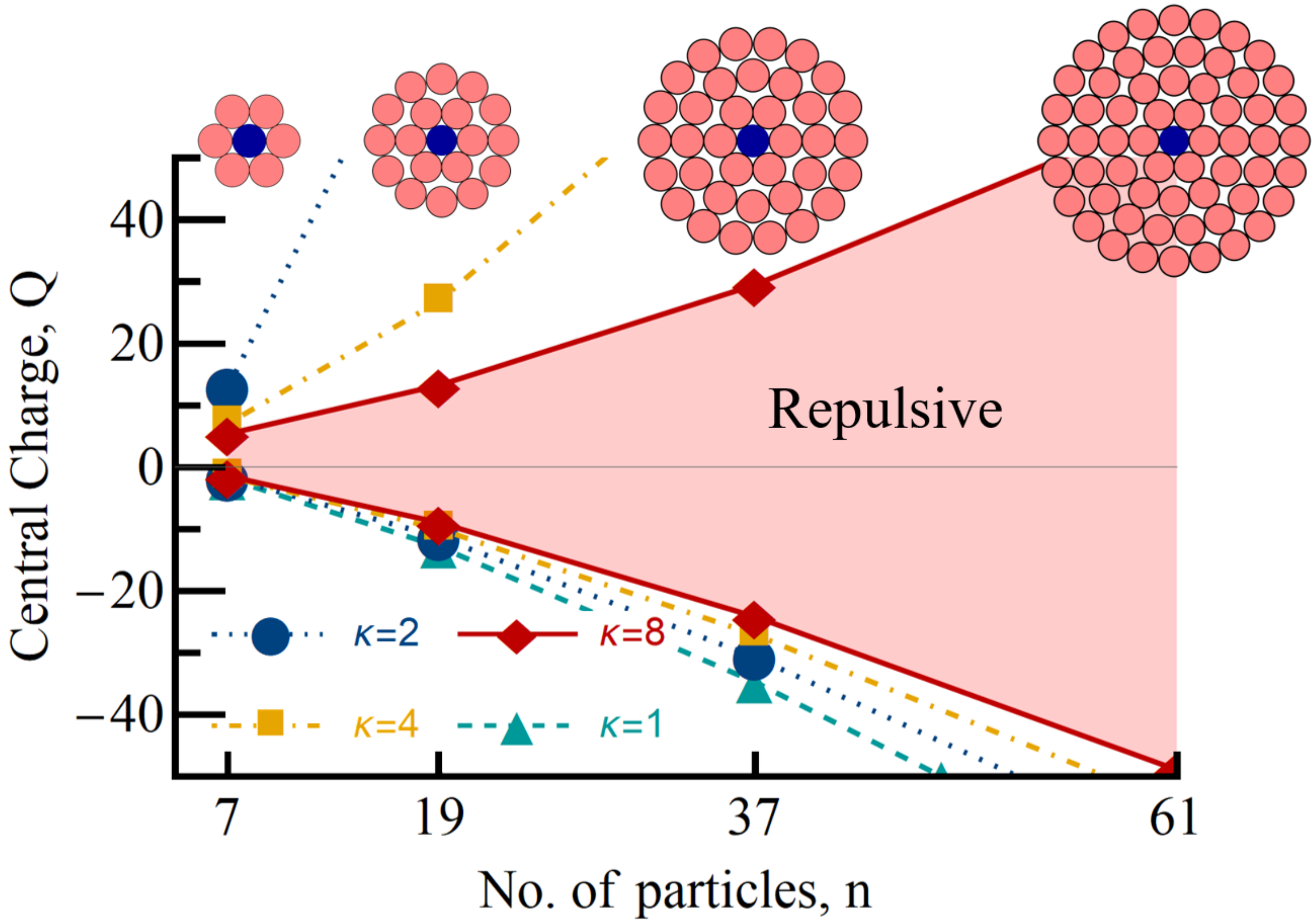}
	\caption{Plot of the limits for mechanical equilibrium, corresponding to the zeros of Eq. \eqref{eq:force_parabola}, as function of the size, $n$, of aggregates for different dielectric constants, $\kappa$ (the corresponding values are close to the values for $n=37$). In the region between the lines, shaded for $ \kappa=8$, particles are repelled from the aggregate.}
	\label{fig:limits}
\end{figure}

Evidently aggregates with eight or fewer particles can be made electrostatically cohesive by the presence of a strongly charged central particle.  To explore larger particle numbers, we simulate successive layers of particles as sketched in the insets to Fig. \ref{fig:limits}.  In the body of this figure, we show the dependence of the limits of cohesion on aggregate size up to $ n=61 $.  The behavior here is similar to that seen for smaller aggregates, except that the magnitude of the central charge by necessity grows with the size of the aggregate. This is easily understood: by symmetry, charges on each layer of surrounding particles must be identical, and so forces between surrounding particles are never attractive \cite{Qin2016}. Thus to maintain a cohesive aggregate as more particles are added, attractive forces due to the central particle must grow to compensate for repulsion of surrounding particles.  As a consequence, charge on the central particle --- and so the separation between the upper and lower curves --- must grow with the size of the aggregate.


\section{Final remarks}
\label{sec:discussion}

In conclusion, we have numerically solved the Poisson equation for aggregates of dielectric particles.  We have confirmed that a central particle can attract large numbers of surrounding dielectric particles --- including particles all of the same sign --- provided that the charge on the central particle is significantly larger than that on its neighbors.  The mechanism at work is simply that the induced polarization on the surrounding particles can overcome Coulombic repulsion.  

We note in closing that the essential feature that distinguishes cohesive from non-cohesive aggregates is therefore the extent of charge heterogeneity present.  Our simulations were by design centrally symmetric, and so cohesive aggregates are characterized by strong variation in radial charge distribution.  This could be viewed as concluding that a single highly charged particle can attract large numbers of weakly charged particles --- and so we predict a highly charged particle can permit planetesimals or industrial agglomerates to grow through the intermediate size range mentioned in the introduction.  

More generally, our central finding is that electrostatic sticking is governed by strong heterogeneity in charge.  We observe that we chose to analyze symmetrically distributed particles, each containing a symmetrically distributed charge, because this defines a well-characterized and unambiguous problem.  Real particles are seldom symmetrically charged \cite{Shinbrot2018} or arranged \cite{Wolf2018} and, to understand real aggregate formation, deeper analysis of both charged particle arrangements and charge distributions within particles is called for.

\begin{acknowledgments}
	We acknowledge financial support from the Luso-American Development Foundation (FLAD), FLAD/NSF, Proj. 273/2016 and the Portuguese Foundation for Science and Technology (FCT) under Contract no. UID/FIS/00618/2013. TS acknowledges support from the NSF DMR award $\sharp$1404792, and CBET award $\sharp$1804286.  We also thank S. Sundaresan, D. Wolf \& G. Wurm for helpful discussions.
\end{acknowledgments}

\bibliography{biblio}

\end{document}